\documentclass[conference]{IEEEtran}
\IEEEoverridecommandlockouts

\usepackage[ruled,linesnumbered]{algorithm2e}

\def\BibTeX{{\rm B\kern-.05em{\sc i\kern-.025em b}\kern-.08em
    T\kern-.1667em\lower.7ex\hbox{E}\kern-.125emX}}
    
\usepackage{CJKutf8}

\usepackage{xcolor}
\usepackage{pgfplots}
\usepackage{tikz}
\usepackage{comment}
\usepackage{makecell}
\usepackage{float}

\usepackage{xurl}

\usepackage{pifont} 
\usepackage{amssymb}
\usepackage{bbding} 
\usepackage{pifont}

\usepackage{makecell}
\usepackage{colortbl} 

\usepackage{fancyhdr} 
\usepackage{multirow}
\usepackage{cite}
\usepackage{amsmath,amssymb,amsfonts}

\usepackage{textcomp}
\usepackage{xcolor}

\usepackage{pgfplots}

\usepackage{graphicx,subfigure}

\definecolor{bblue}{HTML}{4F81BD}
\definecolor{rred}{HTML}{C0504D}
\definecolor{ggreen}{HTML}{9BBB59}
\definecolor{ppurple}{HTML}{9F4C7C}

\usepackage{listings}

\usepackage{blindtext}
\usepackage{etoolbox,xstring,mfirstuc,textcase}
\usepackage{booktabs}
\usepackage{pgfplots}

\usepackage{tabularx,booktabs,makecell,multirow}
\usepackage{enumitem} 

\newcolumntype{L}{>{\arraybackslash}X}

\usepackage{xcolor}
\usepackage{tikz}
\usepackage{pgfplots}

\definecolor{findOptimalPartition}{HTML}{D7191C}
\definecolor{storeClusterComponent}{HTML}{FDAE61}
\definecolor{dbscan}{HTML}{ABDDA4}
\definecolor{constructCluster}{HTML}{2B83BA}

\usepackage[utf8]{inputenc}
\usepackage{listings}
\pgfplotsset{compat=1.18}
\begin{document}

\title{\textsc{FluxLayer}: High-Performance Design for Cross-chain Fragmented Liquidity}
\author{\IEEEauthorblockN{Xin Lao\IEEEauthorrefmark{1},  Shiping Chen\IEEEauthorrefmark{2}, Qin Wang \IEEEauthorrefmark{2}}

\IEEEauthorrefmark{1}  \textit{University of Technology Sydney, Australia} \\
\IEEEauthorrefmark{2} \textit{CSIRO Data61,  Australia}
}

\maketitle
\begin{abstract}
Autonomous Market Makers (AMMs) rely on arbitrage to facilitate passive price updates. Liquidity fragmentation poses a complex challenge across different blockchain networks. 

This paper proposes \textsc{FluxLayer}, a solution to mitigate fragmented liquidity and capture the maximum extractable value (MEV) in a cross-chain environment. FluxLayer is a three-layer framework that integrates a settlement layer, an intent layer, and an under-collateralised leverage lending vault mechanism.  Our evaluation demonstrates that FluxLayer can effectively enhance cross-chain MEV by capturing more arbitrage opportunities, reducing costs, and improving overall liquidity.

\end{abstract}
\smallskip
\begin{IEEEkeywords}
MEV, Cross-chain, Liquidity Fragmentation
\end{IEEEkeywords}

\section{Introduction}

Blockchain software was launched 16 years ago in 2008, following the publication of Bitcoin: A Peer-to-Peer Electronic Cash System \cite{RN56}. Today, it is widely adopted in logistics, finance, and agriculture sectors. As technology advanced, an explosion of new networks ensued with layer 2s (L2s) on Ethereum and Bitcoin, artificial intelligence (AI) and decentralised physical infrastructure networks (DePINs). These new networks all contribute to an increasingly severe liquidity fragmentation problem \cite{RN162}. Today, there are over 200 networks, each with a different ledger system. On each blockchain network, most of the volume of transactions in a given block is arbitrage orders that profit from trading price discrepancies between different exchanges. Since AMM requires arbitrage to update the price, miners are trying to build the most valuable block. Arbitrage is one of the most common forms of maximum extractable value (MEV) \cite{RN61} capture, alongside front-running, back-running, sandwich attacks, liquidation and other strategies. 

Due to the nature of autonomous market makers (AMMs) requiring arbitrage for passive price updates, arbitrage becomes a critical and necessary MEV function. Both Mazor et al.~\cite{RN135} and Sjursen et al.~\cite{RN235} have showcased cross-chain arbitrage MEV, and Danut et al.~\cite{RN239} points out that 87\% of the extraction instances are cyclic cross-chain arbitrage. Moreover, Lehar et al. \cite{lehar2023liquidity} have indicated that liquidity can be fragmented on any blockchain by multiple common factors, including gas price, pool fee, settlement speed, and market maker size. Belchior et al. \cite{RN212} has discussed various solutions to capture cross-chain arbitrage MEV, and more than 100 solutions are on the market. Low-level interoperability protocols offer greater expressiveness and versatility versus asset-specific, chain-specific or application-specific bridges higher up the stack. However, mutually independent solutions perform considerably slower than expected when factoring in the underlying messaging protocols. Most studies focus on arbitrage on a single blockchain, although Mazor tried to fill the gap with cross-chain arbitrage between DEXs \cite{RN135}. Cross-blockchain studies concentrate mainly on general interoperability and message channels \cite{RN212} , but none focus on optimising fragmented liquidity between blockchain networks.

This paper will focus on arbitrage in MEV. Arbitrage occurs on centralised exchanges (CEXs) and decentralised exchanges (DEXs) in various patterns, most notably CEX-to-CEX, CEX-to-DEX and DEX-to-DEX. DEX-to-DEX occurs primarily within a single blockchain, with Ethereum hosting most arbitrage transactions, according to numerous studies~\cite{RN165}\cite{RN159}. Wang et al.~\cite{RN164} points out that Ethereum executed 292,606 cyclic arbitrages over 11 months and extracted more than 138 million USD in revenue. DEX-to-DEX arbitrage is relatively simple and well-understood, however, cross-chain arbitrage is just the beginning. Therefore, this paper will focus on CEX-DEX cross-chain liquidity, also called \textit{non-atomic arbitrage}~\cite{gogol2024layer}. High liquidity fragmentation between the exchanges~\cite{lehar2023liquidity} is the primary cause of CEX-DEX arbitrage opportunities. Arbitrage is critical in balancing liquidity between pools to reduce slippage and give users a better price.

Hereby, we present \textsc{FluxLayer}, an omni-chain intent-centric liquidity framework. FluxLayer utilises intent-centric architecture, Retaking AVS (Active Validator Service), and an under-collateralised leverage lending vault to solve the fractured liquidity problem and provide a faster, cheaper, and easier solution for cross-chain arbitrage. Implementation and simulation results show that FluxLayer can enhance cross-chain MEV by capturing more arbitrage opportunities, reducing costs, and improving overall liquidity.

\section{Methodologies}

The following are methodologies that we have gone through:
\begin{itemize}

    \item Arbitrage definition and Flow - we prove that using the FluxLayer bottom settlement layer and utilising Restaking via AVS consensus to achieve faster finality and help capture the cross-chain MEV at a faster settlement speed.
    \item Simulation with Smart Contract as Custodial - we built a prototype to prove that FluxLayer could work on a smart contract blockchain network like Ethereum.
    \item Simulation with MPC (multi-party computation) wallet as custodial - we built a prototype to prove that FluxLayer could work on any blockchain network, even without a smart contract like the Bitcoin network. 
    \item Non-atomic cross-chain quotes - Both revenue and slippage in a BTC-USDT quote show where the majority of the liquidity of the DEX AMM pool sits. Hence, existing technology that relies on the AMM pool would have limitations on liquidity. 
    \item Non-atomic cross-chain swaps - we were also able to find some random transactions from the quote providers in the block explorer, which include interaction transactions from both DEX and CEX, to showcase the actual cross-chain swap that exists with their revenues and total time.

\end{itemize}

Existing non-atomic cross-chain arbitrage has limitations on pool liquidity, settlement time, high cost and blockchain environments such as EVM (Ethereum Virtual Machine). Whereas FluxLayer could enhance the whole process by speeding up the settlement with AVS Restaking, supporting more blockchain networks (smart contract or MPC non-custodial), enhancing liquidity through intents to connect market makers directly and providing a 10x+ leverage lending vault to help capture more MEV opportunity. 

In this paper, we propose a new framework, FluxLayer, which aims to promptly improve existing solutions in capturing more cross-chain MEV with an infrastructure for omni-chain liquidity. Wang \cite{RN175} classified this type as DApp-based interoperability, and Notland \cite{RN216} classified it as a hybrid of a liquidity and coordination protocol. FluxLayer provides a three-layer design to achieve a faster, cheaper, and easier cross-chain swap experience. Firstly, FluxLayer would introduce the first intent liquidity layer using AVS Restaking, such as EigenLayer \cite{RN186}, for a secure and fast finality. Secondly, it is cheaper; FluxLayer will implement an intent order matching marketplace to reduce cost compared to traditional trusted third parties.

Moreover, being the first to introduce fragment fulfilment for cross-chain swap orders will help with fast fulfilment for large orders and speed up the process of bringing in more competition. Thirdly, FluxLayer would be the first to introduce a cross-chain under-collateralised leverage lending vault, which helps users use under-contribution to leverage capital to create more orders and capture more cross-chain arbitrage opportunities. Ultimately, FluxLayer provides an SDK for the first modular ILaaS (Intent Liquidity as a Service) for easy new DApp/network integration in a permissionless manner. 

\section{Discussion}
\textbf{Flywheel.}
In the intent architecture marketplace, we can see a maker on the left-hand side of the market as a searcher who is good at searching for arbitrage opportunities, but short of capital. And a taker on the right-hand side as a market maker. Then, we can introduce the third important role, the LP (liquidity provider), who has capital and is happy to lend it to generate passive income. Combined with the leverage lending vault above, the maker could borrow assets from the LP in a leveraged way to create more orders; more orders would help takers settle more volumes, generating more revenue that could feed back to the LP and close the loop to become a flywheel. Moreover, LPS would be more likely to provide liquidity in this strategy than the traditional AMM strategy with an impermanent loss \cite{RN209}. This cross-chain arbitrage vault would be safer without impermanent loss as pure arbitrage, but would also bring in higher yields through leverage. Once FluxLayer attracts liquidity onto the platform, it could help streamline upstream liquidity to downstream new blockchain networks. 

\textbf{About MEV.}
There are multiple forms of MEV: liquidation, front/back running, sandwich attack, and arbitrage, which is the most common form. However, arbitrage exists on the cross and single chains; other forms exist mainly in single-chain environments. Arbitrage has been massively captured in a single chain, especially in the Ethereum network. However, with the advancement of technology, more and more networks are booming, which increases the problem of fractured liquidity. Existing solutions have limitations on the blockchain environment, liquidity pool, block finality, high cost, etc. Uniswap Labs, the team behind DEX Uniswap, the number one DEX for the last 6 years, has just launched Unichain \cite{RN229} , solving the liquidity fragmentation problem. However, it relies on the Optimism Superchain \cite{RN230}, which is limited to the EVM environment, and the traditional settlement network architecture requires a lot of work and capital. Therefore, FluxLayer targets all networks and solves the fractured liquidity more efficiently. Moreover, cross-chain is non-atomic and hard to detect, so miners may not know if a transaction is doing cross-chain arbitrage. In other words, cross-chain MEV is less competitive and more benign than single-chain MEV and should be advocated. The current daily cross-chain volume is about \$300m \cite{RN176}; we expect this to be at least tripled shortly, requiring a good architecture design to smooth out the increase in liquidity.

\section{Conclusion}
\label{sec-conclusion}

In this paper, we proposed FluxLayer, a faster, cheaper, and easier cross-chain framework. By integrating with Restaking AVS, we have a quick settlement on a strong security network at a much lower cost with complete flexibility. Moreover, by implementing guaranteed and fragmented fulfilment order types through intent architecture, FluxLayer directly connects market makers to reduce order matching steps and costs and settle large orders. Additionally, FluxLayer uses the under-collateralised leverage lending vault to further enhance liquidity and increase capital efficiency. Hence, the ideal omni-chain liquidity would be a combination of Restaking AVS + intent architect + under-collateralised leverage lending vault. Every element is crucial; missing even one would make the solution incomplete.

\bibliographystyle{unsrt}
\bibliography{bib.bib}

\newpage

\end{document}